\newcommand{\DoPrePrint}{0} % 0 for 2-column submission/review format; 1 for double-spaced, line-numbered preprint
\ifnum\DoPrePrint=1
  \RequirePackage{lineno} 
  \documentclass[aps,prl,preprint,showpacs,superscriptaddress,nofootinbib,linenumbers]{revtex4}

  \usepackage{lineno}
\else
  \documentclass[aps,prl,twocolumn,showpacs,superscriptaddress,nofootinbib]{revtex4-1}
\fi

\usepackage{amsmath}   % need for subequations
\usepackage{graphicx,tikz}  % for figures
\usepackage{xspace}
\usepackage{units}

\usepackage{hyperref}
\usepackage{multirow}

\newcommand{\sizecheck}{0} % 0 to do nothing; 1 to check size
\newcommand{\PRLsupp}{1}   % 0 to do nothing; 1 to put the appendix in a supplement

\ifnum\PRLsupp=0
  
\else
  
\fi

\newif\ifpdf
\ifx\pdfoutput\undefined
   \pdffalse
\else
   \pdfoutput=1
   \pdftrue
\fi
\ifpdf
   \usepackage{graphicx}
   \usepackage{epstopdf}
\else
   \usepackage{graphicx}
\fi

\begin{document}

\ifnum\DoPrePrint=1
\linenumbers
\fi

\title{Measurements of Non-Singlet Moments of the Nucleon Structure Functions and Comparison to Predictions from Lattice QCD for $Q^2 = 4$ $\rm GeV^2$}

%% MANUAL PARTS OF AUTHOR LIST

\newcommand{\deceased}{Deceased}

%% AUTOMATIC LIST (EDITED AS ABOVE)
\newcommand{\ANL}{Argonne National Laboratory, Argonne, IL 60439, USA}
\newcommand{\Basel}{University of  Basel, CH-4056 Basel, Switzerland}
\newcommand{\BU}{Boston University, Boston, MA  02215, USA}
\newcommand{\Newport}{Christopher Newport University, Newport News, VA 23606, USA}  
\newcommand{\Colorado}{University of Colorado, Boulder, CO, USA}
\newcommand{\DFG}{DFG, German Research Foundation, Bonn 51170, Germany}  
\newcommand{\Duke}{Duke University, Dept. of Physics, Box 90305 Durham, NC 27708 }  
\newcommand{\FIU}{Florida International University, Miami, FL 33199, USA} 
\newcommand{\JLAB}{Thomas Jeferson National Accelerator Facility, Newport News, VA 23606, USA}
\newcommand{\Johan}{University of Johannesburg, Auckland Park 2006, Johannesburg, South Africa} 
\newcommand{\JMU}{James Madison University, Harrisonburg, VA  22801, USA} 
\newcommand{\Hampton}{Hampton University, Hampton, VA 23668, USA}
\newcommand{\Houston}{University of Houston, Houston, TX 77004, USA}
\newcommand{\UAE}{Khalifa University of Science and Technology, Abu Dhabi 127788, UAE}
\newcommand{\KEK}{High Energy Accelerator Research Organization (KEK), Tsukuba, Ibaraki 305-0801, Japan}
\newcommand{\LANL}{Los Alamos National Laboratory Los Alamos NM 87545, USA}
\newcommand{\MIT}{Massachusetts Institute of Technology, Cambridge, MA 02139, USA}
\newcommand{\MinD}{Department of Physics, University of Minnesota-Duluth Duluth MN 55812, USA}
\newcommand{\MSU}{Mississippi State University, Mississippi State, MS 39762, USA} 
\newcommand{\NH}{University of New Hampshire,  Durham, NH 03824, USA} 
\newcommand{\Norfolk}{Norfolk State University, Norfolk VA 23504, USA} 
\newcommand{\NW}{Northwestern University, Evanston, IL 60208, USA} 
\newcommand{\NCAT}{North Carolina A\&T  State University, Greensboro NC, 27411, USA} 
\newcommand{\Regina}{University of Regina, Regina, Saskatchewan,  S4S 0A2, Canada }
\newcommand{\Rochester}{Department of Physics and Astronomy, University of Rochester, Rochester, NY  14627, USA}
\newcommand{\USC}{University of Southern California, Los Angeles, CA 90033, USA} 
\newcommand{\SLAC}{Stanford Linear Accelerator Center, Stanford, CA  94025, USA}
\newcommand{\Temple}{Department of Physics, Temple University, Philadelphia, PA 19122, USA} 
\newcommand{\UC}{University of Connecticut, Storrs, CT 06269, USA} 
\newcommand{\CUA}{Catholic University of America, Washington, DC 20064, USA} 
\newcommand{\Tufts}{Physics Department, Tufts University, Medford, MA 02155, USA}
\newcommand{\UVA}{ University of Virginia, Charlottesville, VA 22904, USA}
\newcommand{\Amsterdam}{Vrije Universiteit, Amsterdam, Netherlands}
\newcommand{\Washington}{University of Washington, Seattle, WA 98195, USA}
\newcommand{\WandM}{Department of Physics, College of William \& Mary, Williamsburg, VA 23187, USA}
\newcommand{\Winni}{ University of Winnipeg, Winnipeg, Manitoba R3B 2E9, Canada} 
\newcommand{\Yerevan}{Yerevan Physics Institute, Yerevan, Armenia}
\newcommand{\Zagreb}{University	of Zagreb, Zagreb, Croatia}
\newcommand{\VUnion}{Virginia Union University, Richmond, VA 23220, USA}
\newcommand{\UChicago}{University of Chicago, Chicago, Illinois 60637, USA}

\affiliation{\Hampton}
\affiliation{\ANL} 
\affiliation{\Basel}
\affiliation{\Newport}
\affiliation{\Colorado}
\affiliation{\DFG}  
\affiliation{\Duke}  
\affiliation{\FIU}
\affiliation{\JLAB}
\affiliation{\Johan}
\affiliation{\JMU}
\affiliation{\Houston}
\affiliation{\UAE}
\affiliation{\KEK}
\affiliation{\LANL}
\affiliation{\MIT} 
\affiliation{\MinD} 
\affiliation{\MSU} 
\affiliation{\NH}
\affiliation{\Norfolk}
\affiliation{\NW}
\affiliation{\NCAT}
\affiliation{\Regina}
\affiliation{\Rochester}
\affiliation{\USC}
\affiliation{\Temple}
\affiliation{\UVA}
\affiliation{\Amsterdam}
\affiliation{\WandM}
\affiliation{\Winni}
\affiliation{\Yerevan}
\affiliation{\Zagreb}

\author{I.~Albayrak}
\affiliation{\Hampton}
\affiliation{\CUA}
\author{V.~ Mamyan   }
\affiliation{\UChicago}    
\author{M.~E.~Christy}
\affiliation{\Hampton}
\author{A.~Ahmidouch}
\affiliation{\NCAT}
\author{J.~ Arrington }
\affiliation{\ANL}
\author{A.~ Asaturyan}
\affiliation{\Yerevan }
\author{A.~Bodek}
\affiliation{\Rochester}
\author{ P.~ Bosted }
\affiliation{\WandM}
\author{R. ~Bradford  }
\affiliation{\ANL} 
\author{E. ~Brash  }
\affiliation{\Newport}    
\author{A.~ Bruell   }
\affiliation{\DFG}
\author{C~Butuceanu   }
\affiliation{\Regina}
\author{S.~J.~Coleman}
\affiliation{\WandM}
\author{ M.~Commisso }
\affiliation{\UVA}
\author{ S. ~H.~Connell }
\affiliation{\Johan}
\author{ M.~M.~Dalton}
\affiliation{\UVA}
\author{ S.~Danagoulian}
\affiliation{\NCAT}                    
\author{ A.~Daniel}
\affiliation{\Houston} 
\author{D.~ B.~Day }
\affiliation{\UVA}  
\author{S.~Dhamija }
\affiliation{\FIU} 
\author{J.~ Dunne}
\affiliation{\MSU}
\author{D.~Dutta}
\affiliation{\MSU}
\author{R.~Ent}
\affiliation{\JLAB}
\author{ D.~ Gaskell}
\affiliation{\JLAB}
\author{ A. ~Gasparian}
\affiliation{\NCAT}
\author{R.~Gran}
\affiliation{\MinD} 
\author{T.~Horn}
\affiliation{\CUA}   
\author{Liting Huang}
\affiliation{\Hampton}   
\author{G.~M.~	Huber}
\affiliation{\Regina} 
\author{C.~Jayalath}
\affiliation{\Hampton}
\author{M.~Johnson}
\affiliation{\ANL} \affiliation{\NW}  
\author{M. ~K.~ Jones}
\affiliation{\JLAB}  
\author{N.~Kalantarians}
\affiliation{\VUnion}   
\author{A.~Liyanage}
\affiliation{\Hampton}
\author{C.~E.~Keppel}
\affiliation{\JLAB}
\author{E.~ Kinney}
\affiliation{\Colorado}
\author{Y.~ Li}
\affiliation{\Hampton}   
\author{ S.~Malace }
\affiliation{\Duke}
\author{S.~ Manly}
\affiliation{\Rochester}
\author{P. ~Markowitz}
\affiliation{\FIU}
\author{J.~Maxwell}
\affiliation{\UVA}
\author{N.~N.~Mbianda}
\affiliation{\Johan}
\author{K.~S.~ McFarland}
\affiliation{\Rochester}
\author{M.~ Meziane}
\affiliation{\WandM}
\author{Z.~E.~ Meziani}
\affiliation{\Temple}
\author{G.~B~Mills}
\affiliation{\LANL}
\author{H.~ Mkrtchyan}
\affiliation{\Yerevan}
\author{A.~ Mkrtchyan}
\affiliation{\Yerevan}
\author{J.~ Mulholland}
\affiliation{\UVA}
\author{J.~Nelson}
\affiliation{\WandM}
\author{G.~ Niculescu}
\affiliation{\JMU}
\author{ I.~Niculescu}
\affiliation{\JMU}
\author{ L.~ Pentchev}
\affiliation{\WandM}
\author{ A.~ Puckett}
\affiliation{\UC}
\author{ V.~Punjabi}
\affiliation{\Norfolk}
\author{ I. ~A. ~Qattan}
\affiliation{\UAE}
\author{ P. ~E.~Reimer}
\affiliation{\ANL}
\author{ J. ~Reinhold}
\affiliation{\FIU}
\author{ V.~	M~Rodriguez}
\affiliation{\Houston}
\author{   O.~Rondon-Aramayo}
\affiliation{\UVA}
\author{   M. ~Sakuda}
\affiliation{\KEK}
\author{   W.~ K.~Sakumoto}
\affiliation{\Rochester}
\author{   E.~ Segbefia}
\affiliation{\Hampton}
\author{   T.~ Seva}
\affiliation{\Zagreb}  
\author{I.~Sick}
\affiliation{\Basel}
\author{K.~ Slifer}
\affiliation{\NH}
\author{G.~R~Smith}
\affiliation{\JLAB}
\author{J. ~Steinman}
\affiliation{\Rochester} 
\author{P. ~Solvignon}
\affiliation{\ANL} 
\author{V. ~Tadevosyan}
\affiliation{\Yerevan} 
\author{S.~ Tajima}
\affiliation{\UVA} 
\author{   V. ~Tvaskis}
\affiliation{\Winni}    
\author{ W.~F.~Vulcan}
\affiliation{\JLAB}  
\author{T.~Walton}
\affiliation{\Hampton}   
\author{F.~R~ Wesselmann}
\affiliation{\Norfolk}    
\author{S.~A.~ Wood}
\affiliation{\JLAB}   
\author{Zhihong~Ye}
\affiliation{\Hampton}

%% END AUTOMATIC PART
\collaboration{The E06-009 Collaboration}\ \noaffiliation

\date{\today}


\begin{abstract}
We present extractions of the nucleon non-singlet moments utilizing new precision data on the deuteron $F_2$ structure function at large Bjorken-$x$ determined via the Rosenbluth separation technique at Jefferson Lab Experimental Hall C. These new data are combined with a complementary set of data on the proton previously measured in Hall C at similar kinematics and world data sets on the proton and deuteron at lower $x$ measured at SLAC and CERN.  The new Jefferson Lab data provide coverage of the upper third of the $x$ range, crucial for precision determination of the higher moments. In contrast to previous extractions, these moments have been corrected for nuclear effects in the deuteron using a new global fit to the deuteron and proton data. The obtained experimental moments represent an order of magnitude improvement in precision over previous extractions using high $x$ data. Moreover, recent exciting developments in Lattice QCD calculations provide a first ever comparison of these new experimental results with calculations of moments carried out at the physical pion mass, as well as a new approach which first calculates the quark distributions directly before determining moments.     

\end{abstract}

\ifnum\sizecheck=0  
\maketitle
\fi

In the framework of Quantum Chromodynamics (QCD), the partonic structure of hadrons may be studied through \textit{moments} (or Bjorken $x$ weighted integrals) of the hadron structure functions. The difference of the $u$ and $d$ quark distributions is a flavor non-singlet quantity with the N even (considered in this work) non-singlet moments of these parton distribution functions (PDF) defined as,
\begin{equation}
\label{eq:pdfnsmoment}
\langle x^{N-1} \rangle_{u-d} = \int dxx^{N-1}[u(x)-d(x) + \bar{u}(x)-\bar{d}(x)]. 
\end{equation} 
A successful lattice computation of the nucleon non-singlet moment is a fundamental test of QCD ~\cite{gockeler:lattice2}. Precise Lattice QCD (LQCD) predictions of these moments~\cite{lqcd_1,lqcd_2,lqcd_3,lqcd_4,lqcd_5,lqcd_6,lqcd_7,garret:lattice,Radyuskin_lqcd} are now available. These recent calculations include those that approach the physical pion mass and employ smaller lattice spacing to calculate the moments~\cite{lqcd_1} and those that use an innovative approach to directly calculate the PDFs and from these moments~\cite{garret:lattice}. For a discussion of the connection between PDFs and lattice calculations see \cite{pdf_lattice}. Experimentally, the non-singlet moments can be determined from the difference of proton and neutron $F_2$ moments, obtained from $2F_2^p-F_2^d$, with the deuteron utilized as a proxy for proton plus neutron after correcting for nuclear effects. 

In this letter, we present a precision determination of the non-singlet moments utilizing new measurements of the deuteron $F_2$, in combination with existing proton $F_2$ measurements extracted at a four momentum transfer $Q^2 = 4$ $\rm GeV^2$, to directly confront the lattice results. The extraction of higher moments requires precise data at large $x$, as produced by the new data in the resonance region measured in Jefferson Lab Hall C experiment E06-009. These new measurements facilitate a significant improvement in both precision and accuracy over previous experimental extractions of deuteron and non-singlet nucleon moments~\cite{osipenko:f2moments,niculescu:f2moments}.    

Nucleon structure in terms of quark-gluon momentum distributions is encoded in the unpolarized structure functions $F_1$ and $F_L$, for the exchange of transversely and longitudinally polarized virtual photons respectively,  and $F_2$, which is proportional to $2xF_1 + F_L$.  The total differential cross section can be written in terms of the longitudinal and transverse photoabsorption cross sections as
\begin{equation}
\label{eq:crosssection}
\dfrac{d^2 \sigma}{d \Omega dE^\prime} =\Gamma (\sigma_{T} + \epsilon \sigma_{L})=\Gamma\sigma_r.
\end{equation}
Here $\Gamma=K(\alpha/2\pi^2 Q^2)(E^\prime/E)/(1-\epsilon)$ is the  flux of transverse virtual photons with the total flux $K=\nu (1-x)$ in the Hand convention~\cite{hand:flux}, $\epsilon$ the relative longitudinal flux, $d\Omega$ the differential solid angle and $E$  ($E^\prime$) the energy of the incoming (scattered) electron with four momentum transfer $Q^2$ and energy transfer $\nu=E-E^\prime$. On the right hand side, $\sigma_r$ is called the reduced cross section. Fitting $\sigma_r$ linearly in $\epsilon$ yields $\sigma_L$ as the slope, and  $\sigma_T$ as the intercept. The $F_2$ structure function can then be obtained from
\begin{equation}
 \label{eq:f2}
 F_2(x,Q^2)=\dfrac{K\nu}{4\pi^2\alpha(1+\nu^2/Q^2)}(\sigma_T(x,Q^2)+\sigma_L(x,Q^2)).
\end{equation} 
At leading order, the structure function $F_2$ can be written in terms of the light-cone momentum distribution of partons in the Bjorken limit, ($Q^2\rightarrow \infty$ and at fixed $x$) as
\begin{equation}
\label{eq:f2partons}
F_2=x\sum_i e_i^2 (q_i(x,Q^2)+\overline{q}_i(x,Q^2)).
\end{equation}   
The moments of $F_2$, defined as $\int F_2 x^{N-2} dx$, only receive contributions from operators with spin \textit{N}.  This is not true at finite $Q^2$, where operators with other spins can contribute.  However, Nachtmann~\cite{nachtmann:moments} showed that the contribution to the moments from operators with spin \textit{N} can be projected out by defining moments in terms of the Nachtmann scaling variable $\xi$ as 
\begin{equation}
\label{eq:nmoment}
\begin{split}
&M_2^{(N)}(Q^2) = \int_0^1 dx \frac{\xi^{N+1}}{x^3}\\ &\times(\frac{3+3(N+1)r+N(N+2)r^2}{(N+2)(N+3)}) F_2(x,Q^2),
\end{split}  
\end{equation}
where $N$ is the order of the moment, $\xi=\frac{2x}{1+r}$ is the fraction of the light cone momentum of the struck quark, and $r=\sqrt{1+4M^2x^2/Q^2}$. It is the Nachtmann moments of the data that must be employed for a meaningful comparison to quark distribution moments calculated from LQCD or those determined from perturbative QCD (pQCD) fits.
 	
In the Bjorken limit, structure function moments are independent of $Q^2$ (a phenomenon called \textit{scaling}). At finite $Q^2$, gluon radiative effects, which give rise to scaling violations, and higher twist effects (i.e. interactions between the struck quark and remaining quarks) which give rise to the $Q^2$ dependence of the structure functions, become important. The $Q^2$ dependence of the moments can be studied within the framework of  pQCD, but at lower $Q^2$, pQCD loses its applicability and one must consider finite $Q^2$ effects as well to study the hadronic structure and revert to effective theories or LQCD.   

Current LQCD calculations have focused on non-singlet $u-d$ quantities using moments of the PDFs, which are calculationally simpler because the complicated disconnected diagrams cancel. Experimentally, the integrated non-singlet distribution can be determined from $2p-d$, which is approximately $p-n$, where $p$, $d$ and $n$ denote proton, deuteron and neutron moments respectively. From Eq.~\ref{eq:f2partons}, the non-singlet structure function is 
\begin{equation}
\label{eq:f2ns}
 F_2^p-F_2^n=x\frac{1}{3}(u-d +\bar{u}-\bar{d}) \approx 2F_2^p-F_2^{d} ,
\end{equation}
where $u$ and $d$ are $up$ and $down$ quark distributions, respectively.
Similarly, the non-singlet Nachtmann moments can be determined as $M_2^{NS}=M_2^p - M_2^n \sim 2M_2^p-M_2^{p+n}$, where $M_2^{p+n}$ is obtained from deuteron data as described below.  In the $\overline{MS}$ renormalization scheme, the non-singlet moments of the PDFs, ${\langle x \rangle}_{u-d}$, as calculated in LQCD, which describes the soft, non-perturbative physics, in terms of the non-singlet N=2 moment of the $F_2$ structure function can be written as 
\begin{equation}
\label{eq:f2nsmoment}
{\langle x \rangle}_{u-d}=\dfrac{3}{C_N^v}M_2^{NS}
\end{equation}
where $C_N^v$ are Wilson coefficients which represent the hard, perturbatively calculable coefficient functions. Since PDFs describe non-perturbative behavior, they cannot be directly calculated in perturbative QCD, but they can be calculated using LQCD, or extracted from global fits to a variety of data, for example \cite{CTEQ,MRST,accardi}. 

Although there exist previous deuteron $F_2$ measurements in the nucleon resonance region, those presented in this work are the most precise and accurate determinations to date for several reasons. First, the moments presented here are the first to utilize deuteron and proton $F_2$ values extracted from precision Rosenbluth separations of the structure functions, while previous moment determinations~\cite{osipenko:f2moments} relied on models of the longitudinal contribution. Second, the quasielastic (QE) contribution was precisely determined and then subtracted utilizing the same data set. This is important, because inelastic and quasielastic are treated separately in theory. Third, the deuteron data were corrected for nuclear effects such as Fermi motion, enabling a clean extraction of $p+n$. In all, comparison of these new measurements to the previous $F_2$ moments from \cite{osipenko:f2moments} and \cite{niculescu:f2moments} shows an order of magnitude reduction in the uncertainties.

\begin{figure}[htpb]
\centering
\includegraphics[trim={6.2cm 0.0cm 6.6cm 8.0},clip,width=1.0\columnwidth, angle=0]{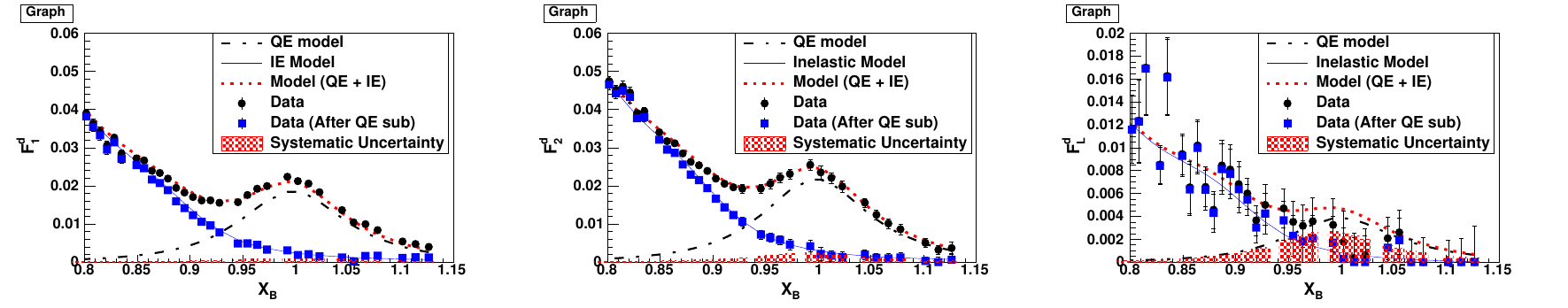}
\caption{(Color online) E06-009 data on deuteron $F_2$ before and after subtraction of the QE contribution. The band at the bottom represents the estimated systematic uncertainty from this procedure. The dot dashed curve is the QE model, the short dashed curve is the total (QE+ inelastic) model and the solid curve is the inelastic model.}
\label{fig:qesub}
\end{figure}

As noted above, inelastic and QE contributions were separated first by removing the latter utilizing the shape of the QE given in \cite{either_qe} with the magnitude determined from the experimental data by scaling up the shape to match the data while the inelastic shape given by a global fit \cite{d2fit} to the available deuteron data. The elastic contribution then was added back at $x = 1$. Figure~\ref{fig:qesub} shows the deuteron structure function $F_2$ in the QE region before and after the QE subtraction. Systematic uncertainties for this subtraction were determined by the following procedure: First, the QE contribution was scaled up and down until the chi-squared value between the data and the fit (QE and inelastic) becomes +1 and -1, and then the difference of the fit from the data was used as the systematic uncertainty. 	
\begin{figure}[htpb]
\centering
\includegraphics[trim={1.0cm 0.0cm 2.0cm 0cm},clip,width=1.0\columnwidth, angle=0]{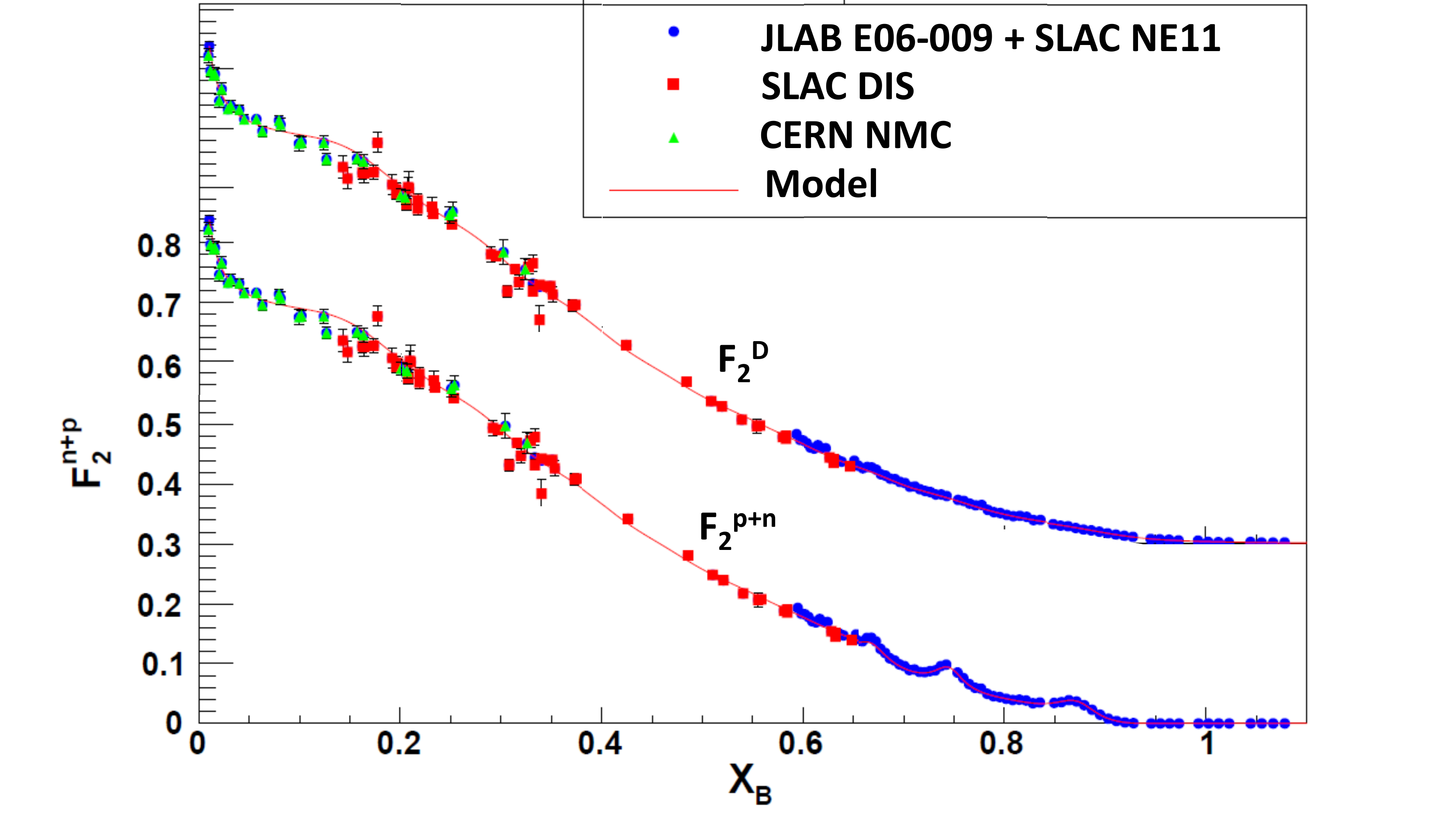}
\caption{(Color online) Top curve: Data on $F_2^d$ at $Q^2 = 4$ $\rm GeV^2$ from SLAC, CERN, and Jefferson Lab experiment E06-009. Top curve is shifted up for comparison to bottom. Bottom curve: Same as top data after Fermi correction.}
\label{fig:worlddata}
\end{figure}
	
Since the deuteron is a bound nucleus and not a pure $p+n$ state, $F_2^d$ was corrected for nuclear effects such as Fermi motion momentum smearing, which washes out the resonant structure, as shown in the top curve of Fig.~\ref{fig:worlddata}.  $F_2^{p+n}$ is obtained from $F_2^{d}$ as $F_2^{p+n}=f(x)F_2^{d}$, where the correction factor is given by
\begin{equation}
\label{eq:corfac}
f(x)=\frac{(F_2^p+F_2^n)^{fit}}{(F_2^d)^{fit}}.
\end{equation}
The global fit to all deuteron data (which determines the neutron) in the resonance region is from \cite{d2fit}. This utilizes the weak binding approximation convolution approach as described in~\cite{owens:model} with the fit to proton data from~\cite{christy:protonfit} as input. 

Fig.~\ref{fig:worlddata} shows existing deuterium data at $Q^2=4$ $\rm GeV^2$ from SLAC~\cite{whitlow:slacdeuteron} and CERN \cite{arneodo:nmcdeuteron} experiments at lower-$x$, as obtained from~\cite{durhamhep:deuteron}, as well as the new precision resonance region data from Jefferson Lab experiment E06-009 at large-$x$,  before nuclear corrections (top), and $p+n$ after the corrections (bottom), where the resonant structure is now quite visible. For $N=2$ moments, the contribution from nuclear corrections is quite small being about $\% 4$ level. The systematic uncertainties due to nuclear corrections are estimated  as the difference of the nominal moments and the nuclear corrected moments utilizing different wave functions based on different nucleon-nucleon potentials (CDBONN, WJC1 and WJC2) \cite{CDBONN,WJC} which represent a spread of behaviors at high momentum in addition to turning on and off the off-shell correction (related to the size of the proton in the nucleus).    

The Nachtmann moments were determined by integrating the combined experimental data shown on the bottom panel in Fig.~\ref{fig:worlddata} using Eq.~\ref{eq:nmoment}. This was accomplished by first dividing the $x$ range into several smaller regions and then fitting the data in each region with fourth order polynomials to provide an interpolating function. The integration range was taken to be from $x =  0.01$ to pion threshold, as there are no deuteron measurements below $x=0.01$. The contribution from $x<0.01$ was estimated to be less than one percent for $N=2$ and negligible for the higher $N$ moments. 

The uncorrelated uncertainties on the moments were determined from a distribution of moments, each calculated from a pseudo-data-set. Individual pseudo-data-sets were generated by sampling about each data point utilizing a Gaussian distribution, with width given by the uncorrelated uncertainty of that data point.        

Sources of correlated systematic uncertainties were due to absolute angle uncertainty, radiative corrections, charge symmetric background subtraction, QE subtraction and Fermi motion corrections. All of these sources of systematic uncertainties were studied in detail. The first three of these systematic uncertainties were studied at the cross section level for the JLab data and propagated to the moments. Details of those studies can be found in~\cite{albayrak:thesis}. The systematic uncertainties due to the QE subtraction and Fermi corrections are discussed earlier in this work. The uncertainties due to the radiative corrections, charge symmetric background subtraction and QE subtraction were found to be small and rather negligible, all being less than 0.1\%, while the systematic uncertainties due to the absolute angle and Fermi corrections provide the largest contributions to the total uncertainty and are given in the tables below. The E06-009 data have been determined from the global fit~\cite{d2fit} to be normalized to better than 
1\% relative to that of SLAC~\cite{whitlow:slacdeuteron} and thus the relative normalization uncertainty contribute negligibly to the uncertainties on the moments. 

Results for the experimental Nachtmann moments of the unpolarized structure function $F_2^{n+p}$ are given in Table~\ref{tab:n_moments_q2_2_3_4} for $Q^2=4$ $\rm GeV^2$. The systematic uncertainties labeled as Sys1, Sys2 and Sys3 are due to the: absolute angle uncertainty, wave function dependence of the Fermi corrections, and off-shell corrections.
\begin{table}[!htbp]
\caption{Experimental Nachtmann moments of $F_2^{n+p}$ (i.e. the deuteron after nuclear corrections) at $Q^2=4$ $\rm GeV^2$. Sys1, Sys2 and Sys3 are the estimated systematic uncertainties due to the absolute angle, wave function and off-shell parameter in the Fermi correction, respectively. The column labeled P2P indicates the quadrature sum o statistical and uncorrelated systematic uncertainties.} 
\centering
\scalebox{0.8}
{
	\begin{tabular}{c c c c c c c c c}
		\hline
		N & $M_2^{p+n}$           & P2P Unc.        & Sys1             & Sys2             & Sys3        \\
		& $(10^{-3})$ & $(10^{-3})$ & $(10^{-3})$ & $(10^{-3})$ & $(10^{-3})$ \\ \hline
		2 & 301.1          & 0.6             & 0.3                 & 0.6             & 3            \\
		4 & 31.4           & 0.1              & 0.2                & 0              & 0.8             \\ 
		6 & 7.8           & 0.02              & 0.1                & 0              & 0.3             \\  \hline
	\end{tabular}
}
\label{tab:n_moments_q2_2_3_4}
\end{table}
Utilizing the previous proton structure function moment determinations given in Table~\ref{tab:moments_proton} from Ref.~\cite{monaghan:privatecomm,pmonaghan:protonfl}, the non-singlet moments of the structure functions were extracted from $M_2^{NS} = 2M_2^p - M_2^{p+n}$. 
\begin{table}[!htbp]
\caption{Experimental Nachtmann moments of proton $F_2$  at $Q^2=3.75$ $\rm GeV^2$ from Ref.~\cite{monaghan:privatecomm,pmonaghan:protonfl} and their corresponding scaled values to $Q^2=4$ $\rm GeV^2$. The uncertainty (Unc.) is the quadrature sum of statistical and uncorrelated systematic uncertainties. } 
\centering
\scalebox{0.9}
{
	\begin{tabular}{ c  c  c  c  c}
		\hline
		N & $M_2^{p}$ ($Q^2=3.75$) & Unc.            & $M_2^{p}$ ($Q^2=4$) & Unc.            \\
		& $(10^{-3})$       & $(10^{-3})$ & $(10^{-3})$    & $(10^{-3})$ \\ \hline
		2 & 173.5                  & 1.8              & 173.0               & 1.9              \\
		4 & 19.9                   & 0.2              & 19.8                & 0.2              \\ 
		6 & 5.1             & 0.1              & 5.1                & 0.1              \\ \hline		
	\end{tabular}
}
\label{tab:moments_proton}
\end{table}

\begin{table}[!htbp]
	\caption{Non-singlet (NS) Nachtmann moments of the unpolarized structure function $F_2$ at $Q^2= 4.0$ $\rm GeV^2$. The moments including the elastic contribution are given in the third column. The fourth column is the experimental $u-d$ non singlet moments obtained as shown in Eq. \ref{eq:f2nsmoment}. Columns five and six are recent LQCD calculations from \cite{garret:lattice} and \cite{lqcd_1}, respectively.} 
	\centering
	\scalebox{0.8}
	{
		\begin{tabular}{ c  c  c  c  c  c}
			\hline
			N & $M_2^{NS}$ & $M_2^{NS}+el.$  & ${\langle x^{N-1} \rangle}_{u-d}^{Exp}$ &  ${\langle x^{N-1} \rangle}_{u-d}^{LQCD1}$&  ${\langle x^{N-1} \rangle}_{u-d}^{LQCD2}$ \\
			  & ($10^{-3})$ & ($10^{-3})$  & ($10^{-3})$ &  ($10^{-3}$)& ($10^{-3}$) \\\hline
			2 & 44.9   (49) & 46.5 (49)  &   138 (14)  &   172 (15)  &   207 (25)   \\
			4 & 8.3   (9) & 9.4 (9)  &   25 (2)   &   24 (3)  &   NA   \\
			6 & 2.4   (3) & 3.0 (3)  &   7.1 (7)   &   NA  &   NA \\ \hline		
		\end{tabular}
	}
	\label{tab:n_m2_non-singlet_moments}
\end{table}

Since the previous proton moments~\cite{monaghan:privatecomm,pmonaghan:protonfl} were evaluated at $Q^2=3.75$ $\rm GeV^2$, they needed to be brought to the common $Q^2$ value of 4 $\rm GeV^2$. This was accomplished utilizing moments calculated from fits to the global data set with the results given in Table \ref{tab:moments_proton}. The uncertainty from this procedure was estimated to be negligible. The non-singlet Nachtmann moments obtained from  $2M^p - M^{p+n}$ are given in Table~\ref{tab:n_m2_non-singlet_moments}, together with the recent LQCD calculations from \cite{garret:lattice} and \cite{lqcd_1}. The experimental $u-d$ moments given in the fourth column is obtained from Eq. \ref{eq:f2nsmoment} as explained below after adding the elastic contribution. It has been suggested~\cite{ji} that the contribution from elastic scattering should be included in to the moments within the operator product expansion, which is utilized by lattice calculations to relate the quark distribution moments to forward nucleon matrix elements of local twist-2 operators. We therefore included this contribution using modern parameterizations of the proton~\cite{JohnProtonFit} and neutron magnetic and elastic form factors, where the neutron form factors were tuned to give better comparisons with the E06-009 data in the quasielastic region. The uncertainty due to this was estimated to be approximately 10\% of the contribution. Wilson coefficients were calculated at next to leading order (NLO) using the prescription given in~\cite{weigl:lattice}:
\begin{equation}
\label{eq:willson}
C_N = C_N^{(0)}+\dfrac{\alpha_s(Q^2)}{4\pi}C_N^{(1)},
\end{equation}    
where $C_N^{(0)} = 1$, $\alpha_s(Q^2)$ is strong force coupling constant calculated with $\Lambda_{QCD}=0.245$ $\rm GeV$ and $C_N^{(1)}$ is the NLO term. The Wilson coefficients were calculated to be $C^{(1)}_2= 1.0104$, $C^{(1)}_4 = 1.142$ and $C^{(1)}_6 = 1.262$ for $N = 2$, $4$ and $6$, respectively. Here, we should also note that current calculations of LQCD include no operators related to resonance production, while integration of physical structure functions to $x=1$ necessitates utilizing resonance region data. Here quark-hadron duality plays an important role and allows direct comparison \cite{wally:duality}. 

The LQCD moments from the QCDSF collaboration~\cite{garret:lattice}, labelled as LQCD1 in 
Table~\ref{tab:n_m2_non-singlet_moments}, result from a novel calculation of full nucleon structure functions on the lattice. The calculation proceeds directly from the virtual Compton amplitude, as outlined in Ref.~\cite{chambers:lattice}, in very much the same way as the moments are extracted from the experimental data, rather than from the leading twist operator matrix element ~\cite{gockeler2_lqcd}. No renormalization is needed. This skirts the issue of renormalization and mixing with operators of higher twist~\cite{martinelli_lqcd}, which impair previous lattice calculations. The moments $\langle x^{N-1}\rangle$ refer to $Q^2=4\,\mbox{GeV}^2$ and are obtained by factoring out the appropriate Wilson coefficient in the $\overline{MS}$ scheme, just like in Eq.~\ref{eq:f2nsmoment}. In contrast, the last column, labeled as LQCD2, is one of the most recent calculations from \cite{lqcd_1} performed at the physical pion mass. LQCD1 provides a great improvement in the agreement with the experimental data while the latter, although performed at the physical pion mass, is systematically higher than the experimental value.

\begin{figure}[htpb]
\centering
\includegraphics[trim={0.0cm 0cm 0.2cm 1.25cm},clip,width=1.1\columnwidth, angle=0]{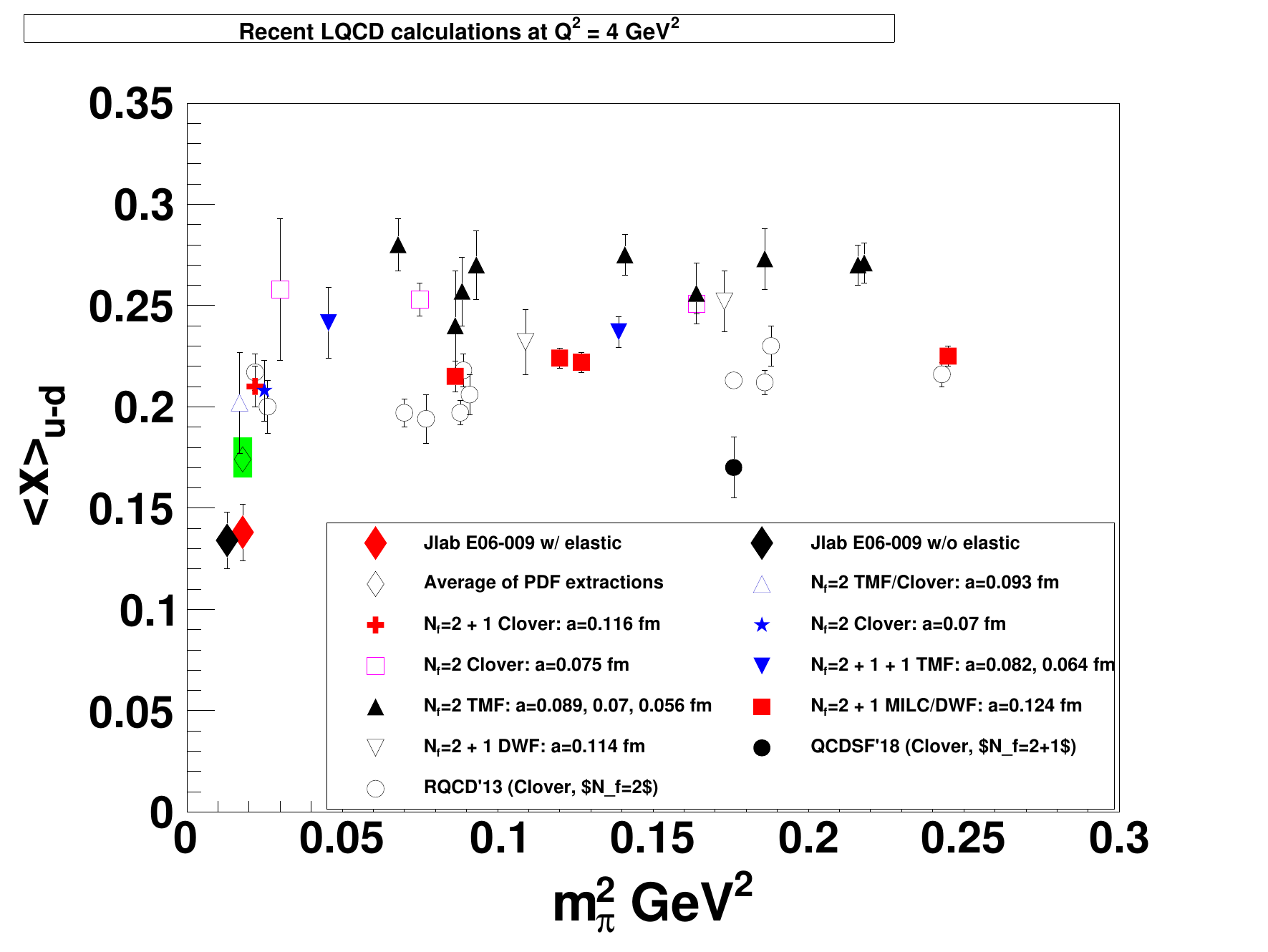}
\caption{(Color online) A collection of recent Lattice QCD calculations for $N=2$ and $Q^2=4$ $\rm GeV^2$ including some (clover) performed at the real pion mass. The experimental result for the N=2 moment is included in the plot for comparison.}
\label{fig:lqcd_all}
\end{figure}

Figure~\ref{fig:lqcd_all} shows a collection of recent Lattice QCD calculations of $N=2$ non-singlet moments  performed at various pion masses at $Q^2=4$ $\rm GeV^2$ together with the experimental results obtained in this analysis. The experimental results are shown with (red diamond) and without (black diamond) the elastic contribution. The one without the elastic contribution (black diamond) is shifted left for clarity purposes. LQCD calculations for twisted mass fermion results (filled triangles, filled inverse triangles and  open triangle) are taken from~\cite{lqcd_1}. Also shown are the results from RBC-UKQCD (filled circle)~\cite{lqcd_2}, LHPC (filled rectangles)~\cite{lqcd_3}, QCDSF/UKQCD (open rectangles)~\cite{lqcd_4}, LHPC (plus marker)~\cite{lqcd_5} and (star)~\cite{lqcd_6}, RQCD (open circle)~\cite{lqcd_8,lqcd_9}. The open diamond is the average of the moments obtained from three different PDF sets determined from pQCD fits~\cite{lqcd_7,PDF1,dbar-ubar2}, with the band indicating the range. Finally the inverted black triangle is the newest calculations from QCDSF \cite{garret:lattice}, which uses a new approach, as explained earlier. 

At the time the that E06-009 experiment was proposed, all LQCD calculations of moments were at large pion mass and disagreed with the values extracted from the available data. Increasing computing power in recent years has made it possible to perform LQCD calculations approaching the real pion mass, thus eliminating the need for calculation at higher values and subsequent extrapolations. The availability of the new precision non-singlet quark moments presented in this work allow several important points can be gleaned from Figure~\ref{fig:lqcd_all}.  First, it is clear that new calculations pushing down to the physical pion mass have not fully resolved the systematic differences with the lower values given by data. Second, new alternative LQCD methods~\cite{garret:lattice}, allowing calculation of the quark distributions directly, give improved agreement with the data and are found to be in agreement within the 1-$\sigma$ uncertainties for not only the $N=2$ moments, but for the $N=4$ moments as well.     
The agreement between the data and PDF extractions may indicate a suppression of higher twists (HT), as recently predicted  in~\cite{glatzmaier}, due to the fact that the cuts applied to remove low $W$ and $Q^2$  from the global data set used to extract these PDFs, the HT is relatively suppressed and likely smaller than the actual resonance region data used here. A systematic study of HT in the moment requires a range in $Q^2$ which can now be performed with our precision data set covering the range $0 < Q^2 < 5$ $\rm GeV^2$. This will be the subject of a forthcoming publication.        

In this paper, we have presented new experimental extractions of the non-singlet $F_2$ structure function moments, as well as non-singlet quark distribution moments with precision many times better than previous extractions. This improved precision is afforded by new precision data on the deuteron $F_2$ structure function from Jefferson Lab experiment E06-009, as well as previous proton and deuteron data from Jefferson Lab, CERN and SLAC. The results have been compared to recent LQCD calculations including those carried out at the real pion mass and those new calculations from QCDSF at a higher pion mass which use a new approach. Although experimental measurements were found to be smaller than the LQCD calculations performed at the real pion mass, they were found to be in far better agreement with those from QCDSF. While there are still problems to overcome in comparing lattice calculations to data, such as residual finite volume effects, renormalization and mixing, these new QCDSF results hint a breakthrough in LQCD calculations after several decades. Improvements are necessary, as high precision data now exist for comparison. The non-singlet quark moments presented in this paper provide a benchmark for LQCD and the study of nucleon structure within QCD and highlight the need to study the differences in LQCD calculations. Additionally, results for higher moments for $N=4$ and $6$ are also presented in this work which can be utilized for confronting future LQCD calculations.  

\ifnum\sizecheck=1
\newpage
{\Large Content after here does not count against size of PRL}
\newpage
\fi

\begin{acknowledgments}

This work was supported in part by NSF grants PHY-1002644 and PHY-1508272 and DOE grants DE-AC02-06CH11357 and DE-FG02-96ER40950. We thank Natural Sciences and Engineering Research Council of Canada (NSERC) for their support.  This material is based upon work supported by the U.S. Department of Energy, Office of Science, Office of Nuclear Physics under contract DE-AC05-06OR23177. We also thank Wally Melnitchouk for useful discussions.

\end{acknowledgments}
\bibliographystyle{apsrev4-1}
\bibliography{moments}

\begin{thebibliography}{42}%
\makeatletter
\providecommand \@ifxundefined [1]{%
 \@ifx{#1\undefined}
}%
\providecommand \@ifnum [1]{%
 \ifnum #1\expandafter \@firstoftwo
 \else \expandafter \@secondoftwo
 \fi
}%
\providecommand \@ifx [1]{%
 \ifx #1\expandafter \@firstoftwo
 \else \expandafter \@secondoftwo
 \fi
}%
\providecommand \natexlab [1]{#1}%
\providecommand \enquote  [1]{``#1''}%
\providecommand \bibnamefont  [1]{#1}%
\providecommand \bibfnamefont [1]{#1}%
\providecommand \citenamefont [1]{#1}%
\providecommand \href@noop [0]{\@secondoftwo}%
\providecommand \href [0]{\begingroup \@sanitize@url \@href}%
\providecommand \@href[1]{\@@startlink{#1}\@@href}%
\providecommand \@@href[1]{\endgroup#1\@@endlink}%
\providecommand \@sanitize@url [0]{\catcode `\\12\catcode `\$12\catcode
  `\&12\catcode `\#12\catcode `\^12\catcode `\_12\catcode `\%12\relax}%
\providecommand \@@startlink[1]{}%
\providecommand \@@endlink[0]{}%
\providecommand \url  [0]{\begingroup\@sanitize@url \@url }%
\providecommand \@url [1]{\endgroup\@href {#1}{\urlprefix }}%
\providecommand \urlprefix  [0]{URL }%
\providecommand \Eprint [0]{\href }%
\providecommand \doibase [0]{http://dx.doi.org/}%
\providecommand \selectlanguage [0]{\@gobble}%
\providecommand \bibinfo  [0]{\@secondoftwo}%
\providecommand \bibfield  [0]{\@secondoftwo}%
\providecommand \translation [1]{[#1]}%
\providecommand \BibitemOpen [0]{}%
\providecommand \bibitemStop [0]{}%
\providecommand \bibitemNoStop [0]{.\EOS\space}%
\providecommand \EOS [0]{\spacefactor3000\relax}%
\providecommand \BibitemShut  [1]{\csname bibitem#1\endcsname}%
\let\auto@bib@innerbib\@empty
%</preamble>
\bibitem [{\citenamefont {Gockeler}\ \emph {et~al.}(2002)\citenamefont
  {Gockeler} \emph {et~al.}}]{gockeler:lattice2}%
  \BibitemOpen
  \bibfield  {author} {\bibinfo {author} {\bibfnamefont {M.}~\bibnamefont
  {Gockeler}} \emph {et~al.},\ }\href@noop {} {\bibfield  {journal} {\bibinfo
  {journal} {Nucl. Phys}\ }\textbf {\bibinfo {volume} {B623}},\ \bibinfo
  {pages} {287} (\bibinfo {year} {2002})}\BibitemShut {NoStop}%
\bibitem [{\citenamefont {Alexandrou}(2015)}]{lqcd_1}%
  \BibitemOpen
  \bibfield  {author} {\bibinfo {author} {\bibfnamefont {C.}~\bibnamefont
  {Alexandrou}},\ }\href@noop {} {\bibfield  {journal} {\bibinfo  {journal}
  {Phys. Rev.}\ }\textbf {\bibinfo {volume} {D92}},\ \bibinfo {pages} {114513}
  (\bibinfo {year} {2015})}\BibitemShut {NoStop}%
\bibitem [{\citenamefont {Aoki}\ \emph {et~al.}(2010)\citenamefont {Aoki} \emph
  {et~al.}}]{lqcd_2}%
  \BibitemOpen
  \bibfield  {author} {\bibinfo {author} {\bibfnamefont {Y.}~\bibnamefont
  {Aoki}} \emph {et~al.},\ }\href@noop {} {\bibfield  {journal} {\bibinfo
  {journal} {Phys. Rev.}\ }\textbf {\bibinfo {volume} {D82}},\ \bibinfo {pages}
  {014501} (\bibinfo {year} {2010})}\BibitemShut {NoStop}%
1003.3387
\bibitem [{\citenamefont {Bratt}\ \emph {et~al.}(2010)\citenamefont {Bratt}
  \emph {et~al.}}]{lqcd_3}%
  \BibitemOpen
  \bibfield  {author} {\bibinfo {author} {\bibfnamefont {J.~D.}\ \bibnamefont
  {Bratt}} \emph {et~al.},\ }\href@noop {} {\bibfield  {journal} {\bibinfo
  {journal} {Phys. Rev.}\ }\textbf {\bibinfo {volume} {D82}},\ \bibinfo {pages}
  {094502} (\bibinfo {year} {2010})}\BibitemShut {NoStop}%
1001.3620
\bibitem [{\citenamefont {Pleiter}\ \emph {et~al.}(2010)\citenamefont {Pleiter}
  \emph {et~al.}}]{lqcd_4}%
  \BibitemOpen
  \bibfield  {author} {\bibinfo {author} {\bibfnamefont {D.}~\bibnamefont
  {Pleiter}} \emph {et~al.},\ }\href@noop {} {\bibfield  {journal} {\bibinfo
  {journal} {PoS}\ }\textbf {\bibinfo {volume} {LATTICE2010}},\ \bibinfo
  {pages} {153} (\bibinfo {year} {2010})}\BibitemShut {NoStop}%
1101.2326
\bibitem [{\citenamefont {Green}\ \emph {et~al.}(2014)\citenamefont {Green}
  \emph {et~al.}}]{lqcd_5}%
  \BibitemOpen
  \bibfield  {author} {\bibinfo {author} {\bibfnamefont {J.}~\bibnamefont
  {Green}} \emph {et~al.},\ }\href@noop {} {\bibfield  {journal} {\bibinfo
  {journal} {Phys. Rev.}\ }\textbf {\bibinfo {volume} {B734}},\ \bibinfo
  {pages} {290} (\bibinfo {year} {2014})}\BibitemShut {NoStop}%
1209.1687
\bibitem [{\citenamefont {Bali}\ \emph
  {et~al.}(2014{\natexlab{a}})\citenamefont {Bali} \emph {et~al.}}]{lqcd_6}%
  \BibitemOpen
  \bibfield  {author} {\bibinfo {author} {\bibfnamefont {G.}~\bibnamefont
  {Bali}} \emph {et~al.},\ }\href@noop {} {\bibfield  {journal} {\bibinfo
  {journal} {Phys. Rev.}\ }\textbf {\bibinfo {volume} {D90}},\ \bibinfo {pages}
  {074510} (\bibinfo {year} {2014}{\natexlab{a}})}\BibitemShut {NoStop}%
1408.6850
\bibitem [{\citenamefont {Alekhin}\ \emph {et~al.}(2012)\citenamefont {Alekhin}
  \emph {et~al.}}]{lqcd_7}%
  \BibitemOpen
  \bibfield  {author} {\bibinfo {author} {\bibfnamefont {S.}~\bibnamefont
  {Alekhin}} \emph {et~al.},\ }\href@noop {} {\bibfield  {journal} {\bibinfo
  {journal} {Phys. Rev.}\ }\textbf {\bibinfo {volume} {D86}},\ \bibinfo {pages}
  {054009} (\bibinfo {year} {2012})}\BibitemShut {NoStop}%
1202.2281
\bibitem [{\citenamefont {Schierholz}()}]{garret:lattice}%
  \BibitemOpen
  \bibfield  {author} {\bibinfo {author} {\bibfnamefont {G.}~\bibnamefont
  {Schierholz}},\ }\href@noop {} {\bibinfo  {journal} {Private Communication}\
  }\BibitemShut {NoStop}%
\bibitem [{\citenamefont {Radyushkin}(2018)}]{Radyuskin_lqcd}%
  \BibitemOpen
\bibfield  {journal} {  }\bibfield  {author} {\bibinfo {author} {\bibfnamefont
  {A.}~\bibnamefont {Radyushkin}},\ }\href {\doibase
  10.1103/PhysRevD.98.014019} {\bibfield  {journal} {\bibinfo  {journal}
  {Phys.Rev. D}\ }\textbf {\bibinfo {volume} {98}},\ \bibinfo {pages} {014019}
  (\bibinfo {year} {2018})}\BibitemShut {NoStop}%
\bibitem [{\citenamefont {Huey-WenLin}\ \emph {et~al.}(2018)\citenamefont
  {Huey-WenLin} \emph {et~al.}}]{pdf_lattice}%
  \BibitemOpen
  \bibfield  {author} {\bibinfo {author} {\bibnamefont {Huey-WenLin}} \emph
  {et~al.},\ }\href {\doibase https://doi.org/10.1016/j.ppnp.2018.01.007}
  {\bibfield  {journal} {\bibinfo  {journal} {Progress in Particle and Nuclear
  Physics}\ }\textbf {\bibinfo {volume} {100}},\ \bibinfo {pages} {107}
  (\bibinfo {year} {2018})}\BibitemShut {NoStop}%
\bibitem [{\citenamefont {Osipenko}\ \emph {et~al.}(2006)\citenamefont
  {Osipenko} \emph {et~al.}}]{osipenko:f2moments}%
  \BibitemOpen
  \bibfield  {author} {\bibinfo {author} {\bibfnamefont {M.}~\bibnamefont
  {Osipenko}} \emph {et~al.},\ }\href {\doibase
  10.1016/j.nuclphysa.2005.11.018} {\bibfield  {journal} {\bibinfo  {journal}
  {Nucl. Phys.}\ }\textbf {\bibinfo {volume} {A766}},\ \bibinfo {pages} {142}
  (\bibinfo {year} {2006})}\BibitemShut {NoStop}%
\bibitem [{\citenamefont {Niculescu}(2006)}]{niculescu:f2moments}%
  \BibitemOpen
  \bibfield  {author} {\bibinfo {author} {\bibfnamefont {M.~I.}\ \bibnamefont
  {Niculescu}},\ }\href {\doibase 10.1103/PhysRevC.73.045206} {\bibfield
  {journal} {\bibinfo  {journal} {Phys. Rev.}\ }\textbf {\bibinfo {volume}
  {C73}},\ \bibinfo {pages} {045206} (\bibinfo {year} {2006})}\BibitemShut
  {NoStop}%
\bibitem [{\citenamefont {Hand}(1963)}]{hand:flux}%
  \BibitemOpen
  \bibfield  {author} {\bibinfo {author} {\bibfnamefont {L.~N.}\ \bibnamefont
  {Hand}},\ }\href@noop {} {\bibfield  {journal} {\bibinfo  {journal} {Phys.
  Rev.}\ }\textbf {\bibinfo {volume} {129}},\ \bibinfo {pages} {1834} (\bibinfo
  {year} {1963})}\BibitemShut {NoStop}%
\bibitem [{\citenamefont {Nachtmann}(1973)}]{nachtmann:moments}%
  \BibitemOpen
  \bibfield  {author} {\bibinfo {author} {\bibfnamefont {O.}~\bibnamefont
  {Nachtmann}},\ }\href@noop {} {\bibfield  {journal} {\bibinfo  {journal}
  {Nucl. Phys.}\ }\textbf {\bibinfo {volume} {B63}},\ \bibinfo {pages} {237}
  (\bibinfo {year} {1973})}\BibitemShut {NoStop}%
\bibitem [{\citenamefont {web page}()}]{CTEQ}%
  \BibitemOpen
  \bibfield  {author} {\bibinfo {author} {\bibfnamefont {C.}~\bibnamefont {web
  page}},\ }\href@noop {} {\bibinfo  {journal} {www.cteq.org}\ }\BibitemShut
  {NoStop}%
\bibitem [{\citenamefont {Martin}\ \emph {et~al.}(2003)\citenamefont {Martin}
  \emph {et~al.}}]{MRST}%
  \BibitemOpen
\bibfield  {journal} {  }\bibfield  {author} {\bibinfo {author} {\bibfnamefont
  {A.~D.}\ \bibnamefont {Martin}} \emph {et~al.},\ }\href@noop {} {\bibfield
  {journal} {\bibinfo  {journal} {Eur. Phys. J}\ }\textbf {\bibinfo {volume} {C
  28}},\ \bibinfo {pages} {455} (\bibinfo {year} {2003})}\BibitemShut {NoStop}%
\bibitem [{\citenamefont {Accardi}\ \emph {et~al.}(2010)\citenamefont {Accardi}
  \emph {et~al.}}]{accardi}%
  \BibitemOpen
  \bibfield  {author} {\bibinfo {author} {\bibfnamefont {A.}~\bibnamefont
  {Accardi}} \emph {et~al.},\ }\href {\doibase 10.1103/PhysRevD.81.034016}
  {\bibfield  {journal} {\bibinfo  {journal} {Physical Review D}\ }\textbf
  {\bibinfo {volume} {81}},\ \bibinfo {pages} {034016} (\bibinfo {year}
  {2010})}\BibitemShut {NoStop}%
\bibitem [{\citenamefont {Ethier}\ \emph {et~al.}(2014)\citenamefont {Ethier},
  \citenamefont {Doshi}, \citenamefont {Malace},\ and\ \citenamefont
  {Melnitchouk}}]{either_qe}%
  \BibitemOpen
  \bibfield  {author} {\bibinfo {author} {\bibfnamefont {J.}~\bibnamefont
  {Ethier}}, \bibinfo {author} {\bibfnamefont {N.}~\bibnamefont {Doshi}},
  \bibinfo {author} {\bibfnamefont {S.}~\bibnamefont {Malace}}, \ and\ \bibinfo
  {author} {\bibfnamefont {W.}~\bibnamefont {Melnitchouk}},\ }\href@noop {}
  {\bibfield  {journal} {\bibinfo  {journal} {Phys.Rev. C}\ }\textbf {\bibinfo
  {volume} {89}},\ \bibinfo {pages} {065203} (\bibinfo {year}
  {2014})}\BibitemShut {NoStop}%
\bibitem [{\citenamefont {Christy}()}]{d2fit}%
  \BibitemOpen
  \bibfield  {author} {\bibinfo {author} {\bibfnamefont {M.~E.}\ \bibnamefont
  {Christy}},\ }\href@noop {} {\bibinfo  {journal} {In Prep.}\ }\BibitemShut
  {NoStop}%
\bibitem [{\citenamefont {Owens}\ \emph {et~al.}(2013)\citenamefont {Owens}
  \emph {et~al.}}]{owens:model}%
  \BibitemOpen
\bibfield  {journal} {  }\bibfield  {author} {\bibinfo {author} {\bibnamefont
  {Owens}} \emph {et~al.},\ }\href@noop {} {\bibfield  {journal} {\bibinfo
  {journal} {Phys. Rev.}\ }\textbf {\bibinfo {volume} {D87}},\ \bibinfo {pages}
  {094012} (\bibinfo {year} {2013})}\BibitemShut {NoStop}%
\bibitem [{\citenamefont {Christy}\ and\ \citenamefont
  {Bosted}(2010)}]{christy:protonfit}%
  \BibitemOpen
  \bibfield  {author} {\bibinfo {author} {\bibfnamefont {M.~E.}\ \bibnamefont
  {Christy}}\ and\ \bibinfo {author} {\bibfnamefont {P.}~\bibnamefont
  {Bosted}},\ }\href@noop {} {\bibfield  {journal} {\bibinfo  {journal} {Phys.
  Rev.}\ }\textbf {\bibinfo {volume} {C81}},\ \bibinfo {pages} {055213}
  (\bibinfo {year} {2010})}\BibitemShut {NoStop}%
\bibitem [{\citenamefont {Whitlow}()}]{whitlow:slacdeuteron}%
  \BibitemOpen
  \bibfield  {author} {\bibinfo {author} {\bibnamefont {Whitlow}},\ }\href@noop
  {} {\bibinfo  {journal} {SLAC-357 (1990) (Ph.D)}\ }\BibitemShut {NoStop}%
\bibitem [{\citenamefont {Arneodo}\ \emph {et~al.}(1997)\citenamefont {Arneodo}
  \emph {et~al.}}]{arneodo:nmcdeuteron}%
  \BibitemOpen
\bibfield  {journal} {  }\bibfield  {author} {\bibinfo {author} {\bibnamefont
  {Arneodo}} \emph {et~al.},\ }\href@noop {} {\bibfield  {journal} {\bibinfo
  {journal} {Nucl. Phys.}\ }\textbf {\bibinfo {volume} {B483}},\ \bibinfo
  {pages} {3} (\bibinfo {year} {1997})}\BibitemShut {NoStop}%
\bibitem [{\citenamefont {Project}()}]{durhamhep:deuteron}%
  \BibitemOpen
  \bibfield  {author} {\bibinfo {author} {\bibfnamefont {D.~H.~D.}\
  \bibnamefont {Project}},\ }\href@noop {} {\bibinfo  {journal}
  {http://durpdg.dur.ac.uk}\ }\BibitemShut {NoStop}%
\bibitem [{\citenamefont {Machleidt}(2001)}]{CDBONN}%
  \BibitemOpen
\bibfield  {journal} {  }\bibfield  {author} {\bibinfo {author} {\bibfnamefont
  {R.}~\bibnamefont {Machleidt}},\ }\href {\doibase 10.1103/PhysRevC.63.024001}
  {\bibfield  {journal} {\bibinfo  {journal} {Phys.Rev. C}\ }\textbf {\bibinfo
  {volume} {63}},\ \bibinfo {pages} {024001} (\bibinfo {year}
  {2001})}\BibitemShut {NoStop}%
\bibitem [{\citenamefont {Gross}\ and\ \citenamefont {Stadler}(2010)}]{WJC}%
  \BibitemOpen
  \bibfield  {author} {\bibinfo {author} {\bibfnamefont {F.}~\bibnamefont
  {Gross}}\ and\ \bibinfo {author} {\bibfnamefont {A.}~\bibnamefont
  {Stadler}},\ }\href {\doibase https://doi.org/10.1103/PhysRevC.82.034004}
  {\bibfield  {journal} {\bibinfo  {journal} {Phys.Rev. C}\ }\textbf {\bibinfo
  {volume} {82}},\ \bibinfo {pages} {034004} (\bibinfo {year}
  {2010})}\BibitemShut {NoStop}%
\bibitem [{\citenamefont {Albayrak}()}]{albayrak:thesis}%
  \BibitemOpen
  \bibfield  {author} {\bibinfo {author} {\bibfnamefont {I.~H.}\ \bibnamefont
  {Albayrak}},\ }\href@noop {} {\bibinfo  {journal} {Jlab HallC dissertation
  (2011) (Ph.D)}\ }\BibitemShut {NoStop}%
\bibitem [{\citenamefont {Monaghan}()}]{monaghan:privatecomm}%
  \BibitemOpen
\bibfield  {journal} {  }\bibfield  {author} {\bibinfo {author} {\bibfnamefont
  {P.}~\bibnamefont {Monaghan}},\ }\href@noop {} {\bibinfo  {journal} {Private
  Communication}\ }\BibitemShut {NoStop}%
\bibitem [{\citenamefont {Monaghan}(2013)}]{pmonaghan:protonfl}%
  \BibitemOpen
\bibfield  {journal} {  }\bibfield  {author} {\bibinfo {author} {\bibfnamefont
  {P.}~\bibnamefont {Monaghan}},\ }\href {\doibase
  10.1103/PhysRevLett.110.152002} {\bibfield  {journal} {\bibinfo  {journal}
  {Phys. Rev. Lett.}\ }\textbf {\bibinfo {volume} {110}},\ \bibinfo {pages}
  {152002} (\bibinfo {year} {2013})}\BibitemShut {NoStop}%
\bibitem [{\citenamefont {Ji}\ and\ \citenamefont {Unrau}(1994)}]{ji}%
  \BibitemOpen
  \bibfield  {author} {\bibinfo {author} {\bibfnamefont {X.-D.}\ \bibnamefont
  {Ji}}\ and\ \bibinfo {author} {\bibfnamefont {P.}~\bibnamefont {Unrau}},\
  }\href {\doibase 10.1016/0370-2693(94)91035-9} {\bibfield  {journal}
  {\bibinfo  {journal} {Phys. Lett.}\ }\textbf {\bibinfo {volume} {B333}},\
  \bibinfo {pages} {228} (\bibinfo {year} {1994})}\BibitemShut {NoStop}%
\bibitem [{\citenamefont {J.~Arrington}\ and\ \citenamefont
  {Tjon}(2007)}]{JohnProtonFit}%
  \BibitemOpen
  \bibfield  {author} {\bibinfo {author} {\bibfnamefont {W.~M.}\ \bibnamefont
  {J.~Arrington}}\ and\ \bibinfo {author} {\bibfnamefont {J.}~\bibnamefont
  {Tjon}},\ }\href@noop {} {\bibfield  {journal} {\bibinfo  {journal}
  {Phys.Rev. C}\ }\textbf {\bibinfo {volume} {76}},\ \bibinfo {pages} {035205}
  (\bibinfo {year} {2007})}\BibitemShut {NoStop}%
\bibitem [{\citenamefont {Weigl}\ and\ \citenamefont
  {Melnitchouk}(1996)}]{weigl:lattice}%
  \BibitemOpen
  \bibfield  {author} {\bibinfo {author} {\bibfnamefont {T.}~\bibnamefont
  {Weigl}}\ and\ \bibinfo {author} {\bibfnamefont {W.}~\bibnamefont
  {Melnitchouk}},\ }\href@noop {} {\bibfield  {journal} {\bibinfo  {journal}
  {Nucl. Physics B}\ }\textbf {\bibinfo {volume} {465}},\ \bibinfo {pages}
  {267} (\bibinfo {year} {1996})}\BibitemShut {NoStop}%
\bibitem [{\citenamefont {Melnitchouk}\ \emph {et~al.}(2005)\citenamefont
  {Melnitchouk} \emph {et~al.}}]{wally:duality}%
  \BibitemOpen
  \bibfield  {author} {\bibinfo {author} {\bibfnamefont {W.}~\bibnamefont
  {Melnitchouk}} \emph {et~al.},\ }\href {\doibase
  10.1016/j.physrep.2004.10.004} {\bibfield  {journal} {\bibinfo  {journal}
  {Physics reports}\ }\textbf {\bibinfo {volume} {406}},\ \bibinfo {pages}
  {127} (\bibinfo {year} {2005})}\BibitemShut {NoStop}%
\bibitem [{\citenamefont {Chambers}\ \emph {et~al.}(2017)\citenamefont
  {Chambers} \emph {et~al.}}]{chambers:lattice}%
  \BibitemOpen
  \bibfield  {author} {\bibinfo {author} {\bibfnamefont {A.~J.}\ \bibnamefont
  {Chambers}} \emph {et~al.},\ }\href@noop {} {\bibfield  {journal} {\bibinfo
  {journal} {Phys. Rev. Lett.}\ }\textbf {\bibinfo {volume} {118}},\ \bibinfo
  {pages} {242001} (\bibinfo {year} {2017})}\BibitemShut {NoStop}%
\bibitem [{\citenamefont {Gockeler}(1996)}]{gockeler2_lqcd}%
  \BibitemOpen
  \bibfield  {author} {\bibinfo {author} {\bibfnamefont {M.}~\bibnamefont
  {Gockeler}},\ }\href@noop {} {\bibfield  {journal} {\bibinfo  {journal}
  {Phys.Rev. D}\ }\textbf {\bibinfo {volume} {53}},\ \bibinfo {pages} {2317}
  (\bibinfo {year} {1996})}\BibitemShut {NoStop}%
\bibitem [{\citenamefont {Martinelli}\ and\ \citenamefont
  {Sachrajda}(1996)}]{martinelli_lqcd}%
  \BibitemOpen
  \bibfield  {author} {\bibinfo {author} {\bibfnamefont {G.}~\bibnamefont
  {Martinelli}}\ and\ \bibinfo {author} {\bibfnamefont {C.~T.}\ \bibnamefont
  {Sachrajda}},\ }\href@noop {} {\bibfield  {journal} {\bibinfo  {journal}
  {Nucl. Phys. B}\ }\textbf {\bibinfo {volume} {478}},\ \bibinfo {pages} {660}
  (\bibinfo {year} {1996})}\BibitemShut {NoStop}%
\bibitem [{\citenamefont {Bali}\ \emph {et~al.}(2013)\citenamefont {Bali} \emph
  {et~al.}}]{lqcd_8}%
  \BibitemOpen
  \bibfield  {author} {\bibinfo {author} {\bibfnamefont {G.~S.}\ \bibnamefont
  {Bali}} \emph {et~al.},\ }\href {\doibase arXiv:1311.7041} {\bibfield
  {journal} {\bibinfo  {journal} {PoS LATTICE 2013}\ ,\ \bibinfo {pages} {290}}
  (\bibinfo {year} {2013})}\BibitemShut {NoStop}%
\bibitem [{\citenamefont {Bali}\ \emph
  {et~al.}(2014{\natexlab{b}})\citenamefont {Bali} \emph {et~al.}}]{lqcd_9}%
  \BibitemOpen
  \bibfield  {author} {\bibinfo {author} {\bibfnamefont {G.~S.}\ \bibnamefont
  {Bali}} \emph {et~al.},\ }\href@noop {} {\bibfield  {journal} {\bibinfo
  {journal} {PoS LATTICE 2014}\ ,\ \bibinfo {pages} {236}} (\bibinfo {year}
  {2014}{\natexlab{b}})}\BibitemShut {NoStop}%
\bibitem [{\citenamefont {Khorramian}\ and\ \citenamefont
  {Tehrani}(2008)}]{PDF1}%
  \BibitemOpen
  \bibfield  {author} {\bibinfo {author} {\bibfnamefont {A.~N.}\ \bibnamefont
  {Khorramian}}\ and\ \bibinfo {author} {\bibfnamefont {S.~A.}\ \bibnamefont
  {Tehrani}},\ }\href@noop {} {\bibfield  {journal} {\bibinfo  {journal} {Phys.
  Rev. D}\ }\textbf {\bibinfo {volume} {78}},\ \bibinfo {pages} {074019}
  (\bibinfo {year} {2008})}\BibitemShut {NoStop}%
\bibitem [{\citenamefont {Blumlein}\ \emph {et~al.}(2007)\citenamefont
  {Blumlein} \emph {et~al.}}]{dbar-ubar2}%
  \BibitemOpen
  \bibfield  {author} {\bibinfo {author} {\bibfnamefont {J.}~\bibnamefont
  {Blumlein}} \emph {et~al.},\ }\href@noop {} {\bibfield  {journal} {\bibinfo
  {journal} {Nucl. Phys.}\ }\textbf {\bibinfo {volume} {B774}},\ \bibinfo
  {pages} {182} (\bibinfo {year} {2007})}\BibitemShut {NoStop}%
\bibitem [{\citenamefont {Glatzmaier}\ \emph {et~al.}(2013)\citenamefont
  {Glatzmaier} \emph {et~al.}}]{glatzmaier}%
  \BibitemOpen
  \bibfield  {author} {\bibinfo {author} {\bibfnamefont {M.~J.}\ \bibnamefont
  {Glatzmaier}} \emph {et~al.},\ }\href {\doibase 10.1103/PhysRevC.88.025202}
  {\bibfield  {journal} {\bibinfo  {journal} {Physical rewiev C}\ }\textbf
  {\bibinfo {volume} {88}},\ \bibinfo {pages} {025202} (\bibinfo {year}
  {2013})}\BibitemShut {NoStop}%
\end{thebibliography}%

\ifnum\PRLsupp=0
\clearpage
\newcommand{\qsq}{\ensuremath{Q^2_{QE}}\xspace}
\renewcommand{\textfraction}{0.05}
\renewcommand{\topfraction}{0.95}
\renewcommand{\bottomfraction}{0.95}
\renewcommand{\floatpagefraction}{0.95}
\renewcommand{\dblfloatpagefraction}{0.95}
\renewcommand{\dbltopfraction}{0.95}
\setcounter{totalnumber}{5}
\setcounter{bottomnumber}{3}
\setcounter{topnumber}{3}
\setcounter{dbltopnumber}{3}

{\normalsize \appendix{Appendix: Supplementary Material}\hfill\vspace*{4ex}}

\fi

\end{document}